\documentclass[11pt,a4paper]{article}
\usepackage[utf8]{inputenc}
\usepackage[T1]{fontenc}
\usepackage{amsmath,amssymb}
\usepackage{graphicx}
\usepackage{booktabs}
\usepackage{subcaption}
\usepackage{siunitx}
\usepackage{xcolor}
\usepackage{hyperref}
\usepackage{xspace}
\usepackage{mhchem}
\usepackage{lineno}
\usepackage[dvipsnames]{xcolor}
\usepackage{xcolor}
\usepackage{float}
\usepackage{authblk}

\newcommand{\dRICH}{dRICH\xspace}
\newcommand{\ePIC}{ePIC\xspace}

\title{Online Data Reduction with Spiking Neural Networks: 
A Temporal-Coincidence Encoder and Distributed SNN for the ePIC dRICH Detector}

\author[1]{Pierpaolo Perticaroli\thanks{Corresponding authors:
  \texttt{pierpaolo.perticaroli@roma1.infn.it},
  \texttt{alessandro.lonardo@roma1.infn.it}}}
\author[2]{Roberto Ammendola}
\author[1]{Andrea Biagioni}
\author[1]{Ottorino Frezza}
\author[1]{Francesca Lo Cicero}
\author[1]{Michele Martinelli}
\author[1]{Pier Stanislao Paolucci}
\author[1]{Elena Pastorelli}
\author[1]{Luca Pontisso}
\author[3,1]{Cristian Rossi}
\author[1]{Francesco Simula}
\author[1]{Piero Vicini}
\author[1]{Alessandro Lonardo}

\affil[1]{Istituto Nazionale di Fisica Nucleare, Sezione di Roma, Rome, Italy}
\affil[2]{Istituto Nazionale di Fisica Nucleare, Sezione di Roma Tor Vergata, Rome, Italy}
\affil[3]{Università Sapienza di Roma, Rome, Italy}
\vspace{-1cm}
\date{}
\begin{document}
\maketitle
\begin{abstract}
The dual-radiator Ring Imaging Cherenkov (dRICH) detector of the \ePIC experiment at the Electron-Ion Collider (EIC) will read out $\sim$\num{320000} silicon photomultiplier (SiPM) channels at a bunch-crossing rate of \SI{100}{\mega\hertz}. The dark count rate (DCR) of the SiPMs is expected to rise up to \SI{300}{\kilo\hertz} per channel over the experiment lifetime, saturating the available output bandwidth and requiring an online data reduction factor of at least five. Most bunch crossings contain only uncorrelated DCR hits, while genuine Cherenkov hits cluster within $\sim$\SI{2}{\nano\second} of the \SI{10}{\nano\second} crossing window: an intrinsically temporal discrimination problem. We present a two-stage online data reduction
pipeline based on spiking neural networks (SNNs). The first stage is a per-photodetection-unit leaky-integrate-and-fire (LIF) \emph{temporal coincidence encoder} that converts raw SiPM hits into a sparse spike stream, achieving more than 90\% data sparsification before any learned classifier is applied. The second stage is a distributed SNN (\num{30} sub-sector networks plus a final aggregation network) deployed on the FELIX-155 Data Aggregation and Manipulation (DAM) boards and on a dedicated Trigger Processor (TP) board, that classifies each bunch crossing as Noise-Only or Signal+Noise. On simulated \ePIC events the system reaches a true positive rate above 94\% at a true negative rate of at least 80\% across the full DCR range; an optional early-exit inference strategy reduces the average classification latency to $\sim$\num{2} algorithmic timesteps at the
cost of a few percentage points on both metrics. A hardware proof-of-concept on an AMD Versal Premium FPGA, integrating
the LIF encoder with the AIGOR multi-core neuromorphic architecture, validates a single sub-sector pipeline at a measured $\sim$\SI{1.7}{\mega\hertz} throughput; ongoing work targets the \SI{100}{\mega\hertz} bunch-crossing rate through an identified set of optimizations of the inference fabric. The methodology may be relevant to other timing-driven detector applications operating at high rate.

\end{abstract}
\section{Introduction}
\label{sec:intro}
Spiking neural networks
(SNNs)~\cite{maass1997networks,gerstner2014neuronal} are a
brain-inspired computing paradigm in which neurons communicate
through discrete spikes in time. They have been repeatedly proposed
for real-time inference on resource-constrained sensor-side systems,
motivated principally by their potential for low-energy, event-driven
computation~\cite{roy2019towards,schuman2022opportunities,edgeSNN2025}:
inference proceeds only when input activity drives neurons to fire,
so the sustained compute cost scales with the actual spike load of
the network rather than being paid in full at every input
sample~\cite{edgeSNN2025}. Their dynamics are intrinsically
temporal, and their hardware footprint can be made small enough for
deployment directly on or near
sensors~\cite{davies2018loihi,edgeSNN2025}. Whether these potential
advantages translate into real-system benefits under stringent
throughput and latency constraints remains an active evaluation
question identified in recent reviews~\cite{edgeSNN2025}.
High-energy and nuclear physics detectors produce massive streams of
finely time-stamped data with strict real-time selection
requirements. Dense neural-network inference on FPGAs has been
established in recent years as an effective approach for online data
reduction and triggering in this
context~\cite{Duarte:2018ite,Aarrestad_2021,Perticaroli:fpgarich}.
The use of SNNs is so far limited to early exploratory
work~\cite{kulkarni2023onsensor,coradin2025neuromorphic}; to our
knowledge no SNN has yet been integrated into a complete HEP DAQ
pipeline.
The dual-radiator Ring Imaging Cherenkov
(\dRICH)~\cite{VALLARINO2024168834} detector of the \ePIC experiment
is a setting in which the question can be addressed concretely. \ePIC
is the first experiment under construction at the Electron-Ion
Collider (EIC), a future facility for high-luminosity electron-proton
and electron-ion collisions~\cite{Abdul_Khalek_2022, ePIC:ref};  the \dRICH
provides charged-hadron identification over a wide momentum range,
from a few \si{GeV}/$c$ up to \SI{50}{GeV}/$c$, reading out
$\sim$\num{320000} silicon photomultiplier (SiPM)
channels~\cite{Rubini:2025ztu} through a multi-stage data-acquisition
chain detailed in Section~\ref{sec:drich}. Over the experiment
lifetime the SiPM dark count rate (DCR) is expected to rise to
$\sim$\SI{300}{\kilo\hertz} per channel under radiation damage, only
partly mitigated by in-situ annealing~\cite{Preghenella:2024epc}; at
that level the front-end raw output exceeds the available egress bandwidth, motivating the development of real-time data reduction strategies. A baseline solution, based on a distributed quantized Multi-Layer Perceptron (MLP) deployed on the readout FPGAs, has
recently been presented~\cite{Rossi:2026}. Two properties of the
\dRICH problem make it of direct interest for SNNs: the signal is
encoded in fine timing structure (Cherenkov hits cluster within
$\sim$\SI{2}{\nano\second} of the \SI{10}{\nano\second} bunch crossing
while DCR is uniform), and the EIC \SI{100}{\mega\hertz}
bunch-crossing rate places SNN inference under severe throughput
pressure. The first property exercises a natural SNN strength; the
second is a stress test of the address-event inference fabric typical
of neuromorphic processors --- specialized hardware that emulates
spiking-neuron dynamics and routes individual spikes as discrete
address events (Address-Event Representation, AER; see
Section~\ref{sec:snns}). The combination makes the \dRICH a clean
setting in which to study both.
We present the design and characterization of an SNN-based online
data reduction system for the \dRICH, structured in two stages: a
per-photo-detection-unit leaky-integrate-and-fire (LIF)
\emph{temporal-coincidence encoder} that achieves $>$90\% data
sparsification before any learned classifier is applied, and a
distributed two-tier SNN classifier deployed on FPGA via the AIGOR
multi-core neuromorphic architecture. We characterize the system in software on
simulated \ePIC events and in hardware execution on AMD
Versal Premium (VPK180) FPGAs. The contributions of this work are an
end-to-end methodology matched to the \dRICH multi-FPGA readout; a
software characterization that meets the $\ge 5\times$ reduction
target with TPR $>$ \num{94}\% across the operational DCR range; and a
hardware-level characterization of the dominant throughput
bottlenecks of an AER-routed FPGA SNN at the boundary of a HEP
front-end. The methodology developed here --- a
temporal-coincidence encoder coupled to a distributed SNN classifier
on FPGA --- provides a template that may be relevant to other
timing-driven detector applications operating at high rate.
The paper is organized as follows. Section~\ref{sec:drich} summarizes
the \dRICH detector, the data reduction requirement, and the simulated
event sample. Section~\ref{sec:snns} motivates the use of SNNs and
reviews the specific throughput challenges they pose.
Section~\ref{sec:arch} describes the LIF encoder, the distributed SNN
topology, and the early-exit inference strategy that together
constitute the technique. Section~\ref{sec:soft} reports the software
characterization of the algorithm on simulated \ePIC events.
Section~\ref{sec:fpga} details the FPGA implementation --- including
the per-PDU encoder pipeline that resolves the front-end data-format
mismatch in a single clock cycle and the deployment of the SNN
classifier on the AIGOR neuromorphic architecture --- and reports the
hardware characterization on a single-sub-sector testbed.
Section~\ref{sec:disc} discusses the main findings and the
optimization path, and Section~\ref{sec:conc} concludes.
\section{The dRICH detector and the data reduction requirement}
\label{sec:drich}
\subsection{Detector and readout chain}

The \dRICH instruments the hadronic endcap of \ePIC with two
Cherenkov radiators (aerogel, refractive index $n \approx 1.02$, and
\ce{C2F6} gas) and a mosaic of spherical mirrors that focus the
emitted Cherenkov photons onto six sectors of SiPM-based
photodetectors, covering a total active area of $\sim$\SI{3}{\square\meter} and comprising
$\sim$\num{320000} readout channels ~\cite{Rubini:2025ztu}. SiPMs were
chosen for their single-photon sensitivity, their tolerance to
magnetic fields of the order of \SI{1}{\tesla}, and their excellent
timing resolution; their main drawback is the radiation-induced
increase of the DCR over the experiment lifetime, mitigated but not
eliminated by periodic high-temperature
annealing~\cite{Preghenella:2024epc}.

The readout follows a hierarchical aggregation scheme structured
around the \emph{Photo Detection Unit} (PDU)~\cite{Cossio:2024nju},
a compact, four-side-buttable assembly comprising a
$16 \times 16$ SiPM matrix (256 channels), four ALCOR-64-based
Front-End Boards (FEBs), and a single FPGA-based Readout Board
(RDO). The ALCOR-64~\cite{Cossio:2024nju} is a 64-channel
mixed-signal ASIC providing per-channel signal amplification,
threshold discrimination, and single-photon time-to-digital
conversion in the \SIrange{25}{50}{\pico\second} range, in a fully
digital, trigger-less architecture; each of its channels reads out a
single SiPM, so the four FEBs of a PDU together cover all
\num{256} channels. The PDU's RDO aggregates the digital streams
from the four FEBs and forwards them over a VTRx+ optical link. The
full detector hosts \num{1248} PDUs.

PDU streams are concentrated by the \emph{Data Aggregation and
Manipulation} (DAM) boards, FELIX-155 cards based on the AMD/Xilinx
Versal Premium architecture~\cite{felix}, with up to \num{48}
optical links per board at \SI{25}{\giga\bit/\second}. The detector
is partitioned into six sectors, each instrumented by five DAMs;
each DAM serves up to \num{42} PDUs of its sub-sector. The
\num{30} DAMs together cover the full detector and forward data to
the \ePIC Echelon-0 processing system through one
\SI{100}{\giga\bit/\second} Ethernet link per board.

\subsection{The bandwidth bottleneck and the reduction target}

In the worst-case DCR scenario of \SI{300}{\kilo\hertz} per channel,
the aggregate raw output of the \dRICH front end is
$\sim$\SI{6.8}{\tera\bit/\second}, with each DAM contributing
$\sim$\SI{226}{\giga\bit/\second}. The available bandwidth to
Echelon-0 is \num{30}$\times$\SI{100}{\giga\bit/\second}, so the
collaboration is targeting a reduction factor of at least five, to
safely bring per-DAM egress inside the budget. The data-reduction
strategy explored in this paper is to identify and discard bunch
crossings that contain only DCR hits, keeping all crossings with any
potential physics content.

\subsection{Why the problem is temporal}

\begin{figure}[htbp]
  \centering
  \includegraphics[width=1.15\linewidth]{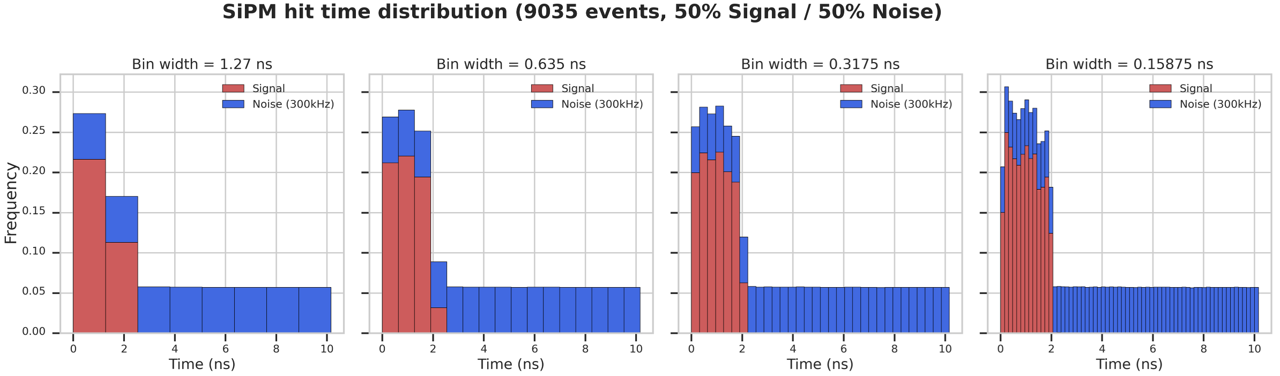}
  \caption{Time distribution of SiPM hits within a single
  \SI{10}{\nano\second} bunch crossing, separated into the
  Cherenkov signal (red) and DCR (blue) components at the
  $\SI{300}{\kilo\hertz}$ operational ceiling. Signal hits cluster
  within $\sim$\SI{2}{\nano\second}, whereas noise is approximately
  uniform across the window.}
  \label{fig:time_distribution}
\end{figure}

The EIC bunch crossing interval is $\tau_{BC} \approx
\SI{10}{\nano\second}$. The reduction system must preserve all bunch
crossings carrying genuine Cherenkov-photon hits, both from physics
signal (deep-inelastic scattering at a peak rate of
$\sim$\SI{500}{\kilo\hertz} across the foreseen beam configurations)
and from beam-induced backgrounds (electron- and hadron-beam-gas
interactions and synchrotron-radiation contributions), with expected rates
spanning from $\sim$\SI{10}{\kilo\hertz} to several \si{\mega\hertz}
depending on beam and accelerator conditions; only bunch crossings
populated exclusively by uncorrelated DCR can be safely discarded.
Within a bunch crossing that contains real Cherenkov hits, the SiPMs
fire within a narrow time window of $\sim$\SI{2}{\nano\second}, set
by the SiPM transit-time spread and by the geometric spread of
photon arrival times for a given Cherenkov ring. DCR hits, by
contrast, are uniformly distributed across the full
\SI{10}{\nano\second} window. This asymmetry, illustrated in
Figure~\ref{fig:time_distribution}, is the lever that any
temporal-coincidence reduction strategy exploits.

The sub-BC time structure is preserved through the readout chain: ALCOR-64 time-tags individual photon hits at the picosecond level, and hits are forwarded to the DAM as individual time-tagged records rather than as per-BC summaries, so sub-BC timing information remains accessible to any classifier operating at the DAM level. The MLP baseline of Rossi~et~al.~\cite{Rossi:2026}, targeting deployment on the same DAM fabric, consumes per-PDU hit counts aggregated over the bunch crossing window and meets the noise-rejection requirement. The fine timing structure of the data and the bunch-crossing rate at which classification must operate make the problem also a natural setting in which to explore model classes whose computational primitives are natively temporal, a direction taken up in the rest of this paper.

\subsection{Simulated event sample and hit characteristics}
\label{sec:dataset}

The system is characterized on simulated \ePIC events produced within
the \ePIC simulation and reconstruction campaigns (Monte~Carlo
generation and reconstruction via \texttt{EICrecon}), with DCR
contributions injected in a dedicated post-processing step. Two event classes are defined.
\emph{Signal+Noise} events combine genuine Cherenkov hits --- from
physics processes (deep-inelastic and semi-inclusive DIS) and
beam-induced backgrounds (electron- and proton-beam--gas interactions,
synchrotron radiation) --- with DCR hits added at the prescribed rate;
\emph{Noise-Only} events contain DCR hits alone. The sample spans the
DCR scan points $25$, $50$, $100$, $150$ and \SI{300}{\kilo\hertz} per
channel, bracketing the operational range expected over the detector
lifetime.

\begin{figure}[htbp]
  \centering
  \includegraphics[width=1.\linewidth]{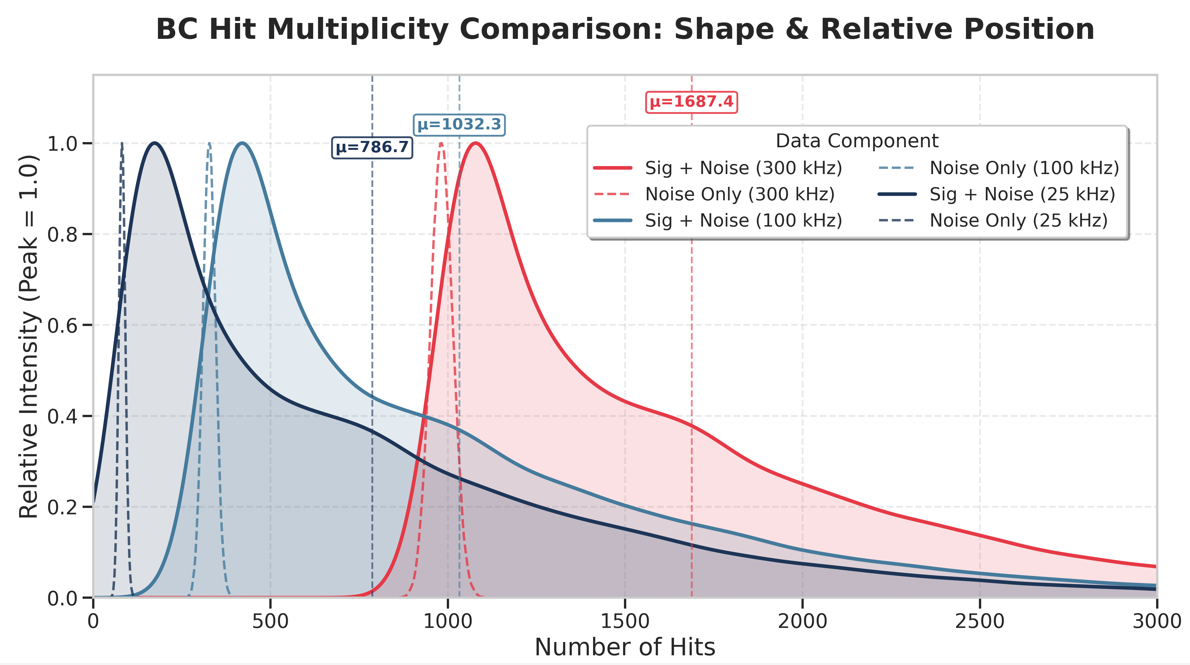}
  \caption{Per-bunch-crossing raw hit multiplicity across the full
  \dRICH at DCR $=25$, $100$ and \SI{300}{\kilo\hertz}. The
  distribution shifts and broadens with DCR, with the
  \SI{300}{\kilo\hertz} case dominated by the $\sim$\num{1000}-hit
  uncorrelated background.}
  \label{fig:hit_multiplicity}
\end{figure}

The two classes differ both in time structure
(Fig.~\ref{fig:time_distribution}) and in hit multiplicity. At the
\SI{300}{\kilo\hertz} ceiling the average raw occupancy is
$\sim$\num{1687} hits per bunch crossing across the full \dRICH, of
which $\sim$\num{1000} are uncorrelated DCR;
Figure~\ref{fig:hit_multiplicity} shows how the per-BC multiplicity
distribution scales with DCR. These two features --- the
$\sim$\SI{2}{\nano\second} Cherenkov clustering and the DCR-dominated
occupancy --- are precisely what the encoder of
Section~\ref{sec:encoder} is built to exploit. Training-specific
details (event counts and partitioning) are deferred to
Section~\ref{sec:soft}.

\section{Spiking neural networks for online data reduction}
\label{sec:snns}

\subsection{Motivation}

Spiking neural networks process information through discrete
spikes that occur at specific points in time. Each neuron maintains
an internal state variable $V(t)$ (the \emph{membrane potential})
that integrates inputs over time and decays (\emph{leaks}) between
inputs; when $V(t)$ exceeds a threshold $\theta$, the neuron emits a
spike, $V(t)$ is reset, and the spike is propagated downstream. The
leaky-integrate-and-fire (LIF) model captures these dynamics in a
single equation:
\begin{equation}
  V(t+\Delta t) \;=\; \alpha\, V(t) \;+\; \sum_j w_j\, x_j(t),
  \qquad
  \text{spike if } V(t+\Delta t) > \theta,
  \label{eq:LIF}
\end{equation}
where $\alpha \in [0,1)$ is the leak factor, $w_j$ the synaptic
weights, and $x_j(t)$ the (binary) spike inputs at timestep
$t$~\cite{gerstner2014neuronal}.

For our problem, three properties of SNNs are
particularly attractive:
\emph{native temporal encoding} --- information lives in spike timing
rather than in features aggregated over fixed time windows, so the
\SI{2}{\nano\second} burst structure of Cherenkov hits can be exploited
without explicit feature engineering;
\emph{event-driven sparsity} --- compute is performed only when spikes
are present, so the cost of processing the bunch crossings dominated
by uncorrelated DCR alone can in principle be very small;
and \emph{compact, hardware-friendly inference} --- the LIF neuron is a
small, fixed piece of arithmetic that maps well to digital
hardware, in principle amenable to deployment closer to the front-end
electronics than the DAM-side implementation of the present work.
\subsection{Design constraints for SNN inference in a DAQ pipeline}

Realizing these advantages in a real DAQ pipeline requires addressing
three design constraints, each of which informs the architectural
choices described in Section~\ref{sec:arch}.
\paragraph{AER spike routing under DCR load.} Neuromorphic
hardware typically routes spikes by Address-Event Representation
(AER)~\cite{AER:ref}: each spike traverses a shared bus as a small
packet carrying the originating neuron's ID and timestamp. While
Cherenkov signals are sparse in space, the persistent DCR background
creates a steady flood of hits that, if not pre-filtered, would
congest the AER fabric on every bunch crossing.
\paragraph{Time-resolving packed hit words at the AER boundary.} When SNN
inference is routed through an AER bus, as is the case for the AIGOR
neuromorphic architecture used in this work, packed hit words from the
front-end have to be \emph{time-resolved} into individual spike events
before being injected into the routing fabric. A naive serial
implementation would cost one clock cycle per hit in a word, consuming
the per-PDU latency budget and reducing throughput.
Section~\ref{sec:encoder_fw} describes the parallel LIF cascade we
adopt to collapse this time-resolution into a single clock cycle under
the single-spike-per-BC assumption.
\paragraph{Timestep synchronization.} Digital SNN accelerators with
timestep-based dynamics must mark the advance of algorithmic time
explicitly on their communication fabric: each core signals the
completion of every timestep, and downstream cores wait for this
synchronization information before advancing, so that spike ordering
is reproducible. On sparse inputs --- precisely the regime the encoder
of Section~\ref{sec:arch} creates --- many timesteps carry few or no
spikes, and the synchronization traffic can come to dominate the AER
transactions. As discussed in Section~\ref{sec:hw}, this is the
dominant throughput cost of the present implementation. Fusing the
synchronization information into the last spike word of a timestep
removes the dedicated transactions, but the cost is structural to
timestep-driven execution: addressing it fully requires evolving the
execution model toward event-driven dynamics.

Each of these constraints is addressed within the design space we
explore: the LIF encoder (Section~\ref{sec:arch}) reduces the rate at
which spikes ever reach the AER fabric, relaxing both the
time-resolution and bandwidth pressures, and the synchronization
cost, structural to timestep-driven execution, is the primary target
of the ongoing evolution of the inference fabric
(Section~\ref{sec:hw}).

\subsection{Related work}

The use of neuromorphic hardware for HEP triggering and front-end
filtering has been explored in preliminary form in
\cite{kulkarni2023onsensor} for on-sensor data filtering and in
\cite{coradin2025neuromorphic} for unsupervised particle tracking.
On the FPGA side, dense neural-network inference at the MHz scale has been demonstrated for trigger and online-selection systems in a number of HEP
contexts~\cite{Duarte:2018ite,Aarrestad_2021,Govorkova2022,CMS:P2L1T,Aaij2020Allen,Perticaroli:fpgarich};
the MLP-based \dRICH data reduction of~\cite{Rossi:2026} addresses the same task on the same readout fabric.
The combination of (i) an
explicit temporal-coincidence encoder, (ii) a distributed SNN
classifier matched to a sectorized FPGA-based readout, and (iii) a
deployment validated on the same hardware fabric that hosts the
existing DAQ is, to our knowledge, new.

\section{System architecture: LIF encoder and distributed SNN}
\label{sec:arch}

\subsection{Two-stage design}

\begin{figure}[htbp]
  \centering
  \includegraphics[width=1\linewidth]{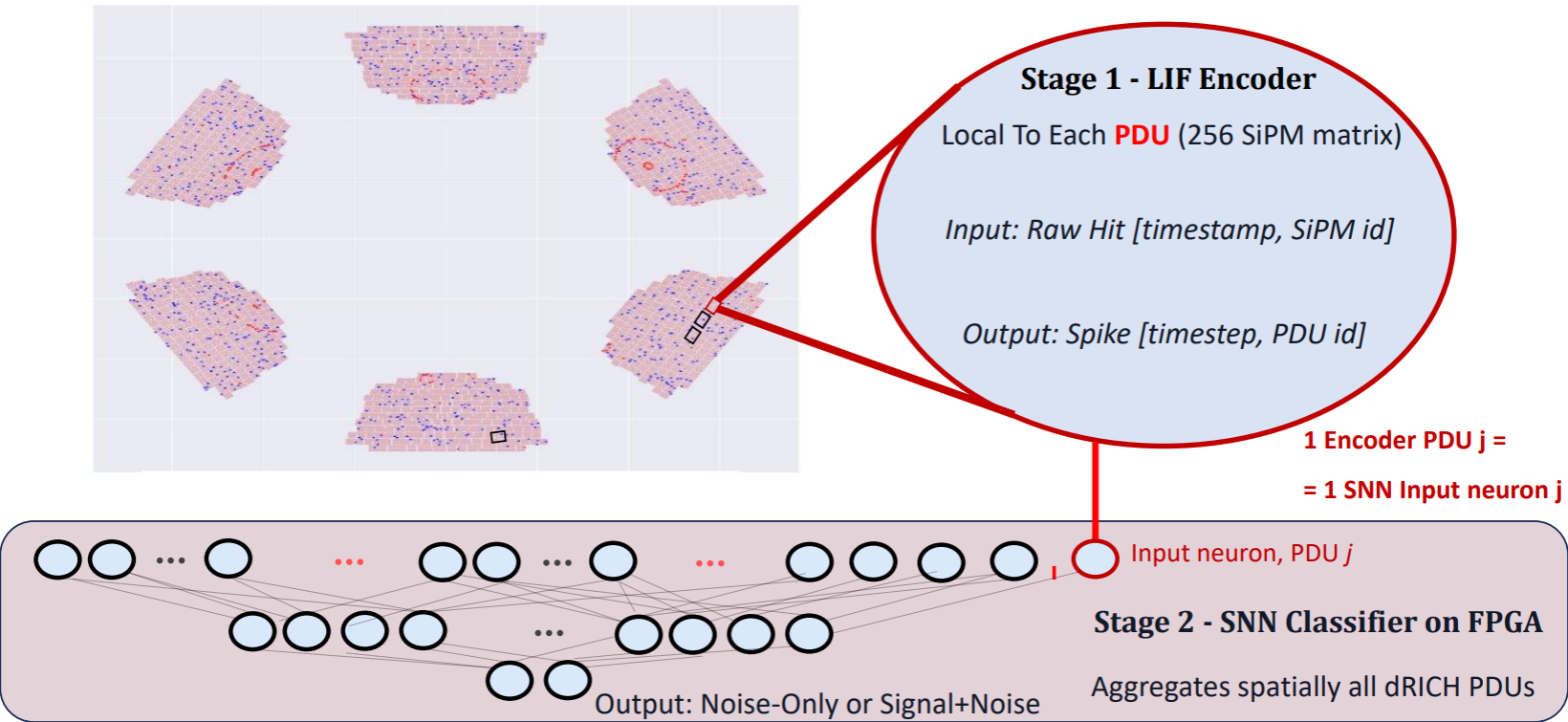}
  \caption{Two-stage pipeline. Stage~1: per-PDU LIF temporal-coincidence
  encoder, local to each photo-detection unit (256 SiPM inputs $\to$
  one LIF input neuron). Stage~2: distributed SNN classifier, with
  \num{30} sub-sector sub-networks (one per DAM) and one aggregation
  sub-network on the Trigger Processor, producing the Signal+Noise vs
  Noise-Only decision.}
  \label{fig:system_overview}
\end{figure}

The proposed pipeline is structured in two stages
(Figure~\ref{fig:system_overview}).
\textbf{Stage~1 --- Encoder.} For each of the \num{1248} PDUs, a single
LIF neuron whose 256 inputs are the SiPM channels of that PDU
integrates the incoming hits in time and emits at most one spike per
bunch crossing, retaining only the first threshold crossing
(\textit{single-spike-per-BC}). The output of Stage~1 is therefore a
sparse spike stream of dimensionality \num{1248} (one binary line per
PDU per BC), accompanied by sub-BC timing information.
\textbf{Stage~2 --- Classifier.} A multi-layer SNN distributed over the
entire \dRICH consumes the encoder output and, after a small number of
timesteps, classifies the bunch crossing as Signal+Noise or
Noise-Only. The decision is fanned out to the DAMs via the \ePIC Global
Timing Unit (GTU), the detector-wide clock-and-command distribution
system, which gates the forwarding of buffered RDO data to Echelon-0
--- the first off-detector tier of the \ePIC streaming-readout chain,
where surviving bunch crossings are received over the
$30\times$\SI{100}{\giga\bit/\second} Ethernet links.

\subsection{The LIF encoder as a temporal coincidence detector}
\label{sec:encoder}

The encoder is, conceptually, the heart of the technique. Each PDU
hosts a single LIF neuron whose 256 input weights are kept equal
(uniform $w$) --- the encoder does not need to discriminate among SiPM
positions within a PDU, only among temporal patterns. With this
choice, Equation~\ref{eq:LIF} reduces to
\begin{equation}
  V(t+\Delta t)
  \;=\; \alpha\, V(t) \;+\; w \cdot n_{\rm hit}(t),
  \qquad
  \text{spike if } V(t+\Delta t) > \theta,
  \label{eq:encoder}
\end{equation}
where $n_{\rm hit}(t)$ is the number of SiPM channels of the PDU that
fired within the timestep $\Delta t$. The dynamics of the encoder
under the two input classes are easy to read off:
a burst of Cherenkov hits, all arriving within a single
$\Delta t$ or two, rapidly drives $V$ above $\theta$ and produces a
spike;
a trickle of DCR hits, distributed uniformly over the bunch crossing,
accumulates more slowly than the leak depletes $V$, and the
neuron remains silent.

The encoder is fully specified by four hyperparameters: the timestep
size $\Delta t$, the input weight $w$ (which we keep at unity and absorb
into $\theta$), the leak factor $\alpha$, and the threshold $\theta$.
For a hardware-friendly implementation the leak is realized as a
fixed-point right shift: either as a true geometric leak with
$\alpha = 1-2^{-k}$, or, in the most aggressive configuration, as a
plain division of $V$ by a power of two each timestep. The membrane is
held in low-width fixed-point or integer arithmetic. The configuration
space therefore spans a continuum from a graded temporal integrator
down to, in the limiting case of a single-bit membrane with
$\theta = 2$ and full per-timestep reset, a pure same-bin
\emph{coincidence detector} that fires when two or more channels of a
PDU hit within one time bin. Both ends of this continuum are
multiplier-free: the graded leak costs only a shift and an add per
timestep on a few-bit register, and the coincidence limit removes even
that, collapsing the membrane update to a single-bit same-bin test. This
last limit is the configuration we deploy in hardware
(Section~\ref{sec:encoder_fw}); the configuration scan of
Section~\ref{sec:soft} shows it stays within a small accuracy margin of
the graded-leak variants, so the cheapest design point is also the one
we adopt.

Crucially, the encoder is itself a data reduction stage. At the
\SI{300}{\kilo\hertz} operational ceiling the raw occupancy is
dominated by DCR (Section~\ref{sec:dataset}: $\sim$\num{1000} of
$\sim$\num{1687} hits per BC, Figure~\ref{fig:hit_multiplicity}). After
encoding with a well-chosen $(\Delta t, k, \theta)$, the total number
of spikes per bunch crossing across the full detector drops to
$\sim$\num{60}--\num{120}, with a peak load on any single sub-sector of
$\sim$\num{14} spikes per BC (Figure~\ref{fig:encoder_spikeload}). The
encoder by itself therefore performs $>$90\% sparsification, leaving the
downstream SNN with a budget compatible with AER-based routing.


\subsection{The distributed SNN classifier}
\label{sec:classifier}

\begin{figure}[htbp]
  \centering

  \begin{subfigure}{\linewidth}
    \centering
    \includegraphics[width=1.05\linewidth]{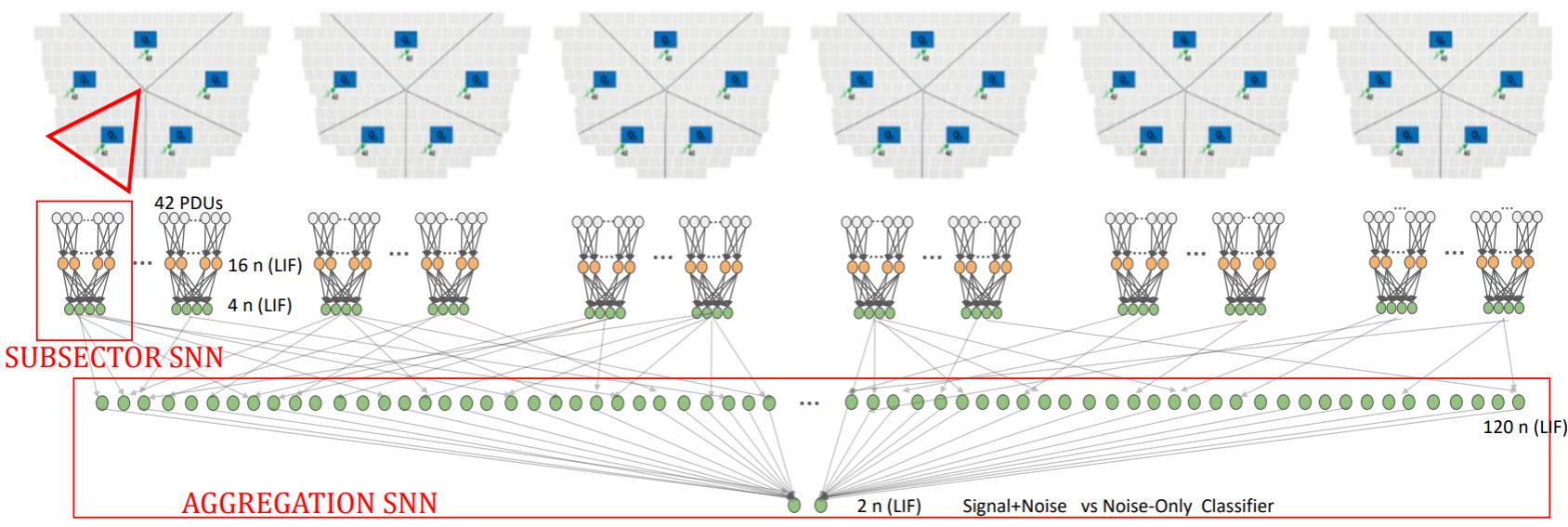}
    \caption{}
    \label{fig:distributed_snn_a}
  \end{subfigure}

  \vspace{0.8em}

  \begin{subfigure}{\linewidth}
    \centering
    \includegraphics[width=1.05\linewidth]{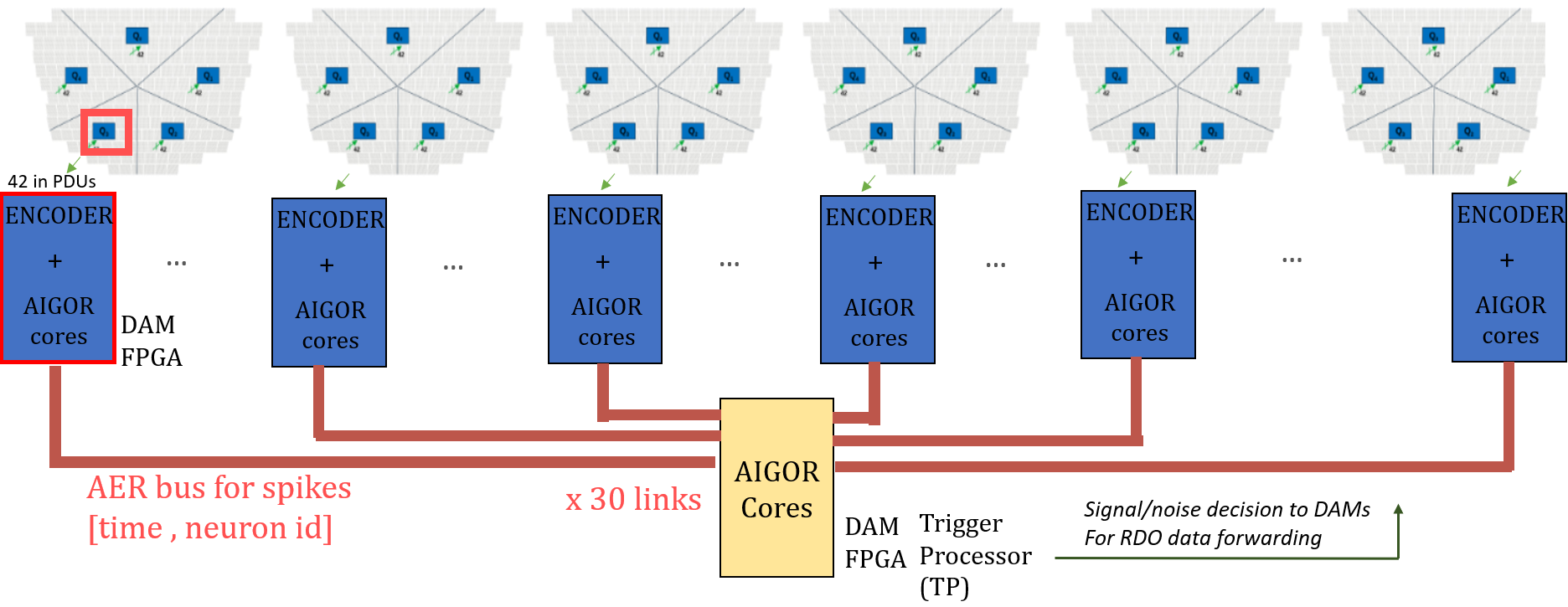}
    \caption{}
    \label{fig:distributed_snn_b}
  \end{subfigure}

  \caption{Two views of the distributed two-tier SNN. (a)~Algorithmic flow: each Sub-sector SNN consumes the \num{42} encoder outputs of one DAM through a \num{16}-neuron and a
  \num{4}-neuron LIF layer; the Aggregation SNN concatenates the \num{30}$\times$\num{4} local features into a \num{120}-neuron hidden layer and a \num{2}-neuron output that emits the Signal+Noise vs Noise-Only verdict. (b)~The same structure mapped onto the hardware: each DAM FPGA hosts the encoder and the AIGOR cores of its Sub-sector SNN and streams the local features over a dedicated AER link ([time, neuron id]) to the Trigger Processor, a further DAM board hosting the Aggregation SNN cores; the trigger decision is returned to the \num{30} DAMs, where it gates the forwarding of the buffered RDO data.}
  \label{fig:distributed_snn}
\end{figure}

The classifier topology mirrors the physical readout
(Figure~\ref{fig:distributed_snn}). Each sub-sector hosts a
\emph{Sub-sector SNN} whose input layer consists of the \num{42} LIF
encoder outputs of that sub-sector, followed by one hidden LIF layer
of \num{16} neurons and one further LIF layer of \num{4} neurons.
The 4-dimensional output of each Sub-sector SNN constitutes a compact
local feature vector. The 30 Sub-sector SNNs feed an
\emph{Aggregation SNN}, hosted on the Trigger Processor, whose input
is the concatenation of the \num{120} local features ($30 \times 4$);
one hidden LIF layer of \num{120} neurons drives a final \num{2}-neuron
LIF output that produces the binary classification.

The topology reflects two deliberate constraints. First, the locality
of the Sub-sector SNN is exact: each DAM sees only its own \num{42}
PDUs, with no inter-DAM communication during sub-sector inference, so
the 30 inference paths can run in parallel at high throughput; the inter-FPGA fabric
carries only the four-dimensional feature vectors to the Trigger Processor. 

Second, the encoder and the classifier layers both derive from the LIF
model of Eq.~\ref{eq:LIF}, but they are not the same block. The
classifier layers run on AIGOR with trained per-synapse
weights and may emit multiple spikes per timestep, whereas the encoder
is a specialized, heavily approximated LIF --- uniform weights, a
single-bit membrane, a shift-based leak, and a one-spike-per-BC policy
--- implemented separately in dedicated firmware
(Section~\ref{sec:encoder_fw}). Sharing the neuron abstraction keeps
the pipeline conceptually uniform; specializing the encoder is what
makes the front-end stage cheap enough to run at rate.

\subsection{Early-exit inference}
\label{sec:earlyexit}

SNN inference is iterative: the network state advances each timestep
and a verdict is normally read out after a fixed number $T_{\max}$ of
steps. The two classes are strongly asymmetric in output activity ---
Signal+Noise events drive the Signal output neuron to fire within the
first few timesteps, while Noise-Only events leave it largely silent
--- and this asymmetry can be exploited to halt inference as soon as
the verdict is unambiguous. In the simulated sample used throughout this work
(Section~\ref{sec:dataset}) the Cherenkov hits arrive early within the
bunch crossing, which makes the effect especially pronounced.

We adopt the following \emph{early-exit} policy. At each timestep $t$
we accumulate the output spike counts of both classes, $n_{\rm sig}(t)$
and $n_{\rm noise}(t)$. The first class whose count reaches a
configurable threshold $ET$ wins and inference halts, with the Signal
class taking priority when both cross in the same step. If $T_{\max}$
timesteps elapse without either class reaching $ET$, the verdict falls
back to a rate-coding decision on the accumulated counts (Signal if
$n_{\rm sig}/(n_{\rm sig}+n_{\rm noise})$ exceeds a fixed ratio,
Noise-Only otherwise). This trades a small loss in classification
performance (a few percent on TPR/TNR, Section~\ref{sec:soft}) for a
substantial reduction in average latency: at $ET = 1$, the average
Signal classification finishes in $\sim$\num{2} timesteps, and the
99th-percentile latency is $\sim$\num{5} timesteps.

\section{Software characterization}
\label{sec:soft}

\subsection{Training sample and partitioning}

The event classes and hit characteristics of the simulated sample are
described in Section~\ref{sec:dataset}. For training, events with zero
\dRICH hits at reconstruction level are removed from the Signal+Noise
class, and a balanced set of \num{80000} events is assembled across the
two classes. The data are split 80\%/10\%/10\% into training,
validation, and test partitions.

\subsection{Training and operating-point selection}

The classifier is trained by surrogate-gradient backpropagation through
time~\cite{neftci2019surrogate} in snnTorch~\cite{snntorch}, with the non-differentiable
spike replaced by a smooth surrogate on the backward pass and
cross-entropy on the final-timestep output of the Aggregation SNN as
the loss. The encoder is \emph{not} co-trained by gradient descent: its
hyperparameters $(\Delta t, \theta, k)$ define a small discrete
configuration space, and each configuration is evaluated end-to-end by
training the downstream classifier on its spike output and measuring
the resulting TPR/TNR and inference latency (early exit strategy). The operating point is then selected from this scan (Section~\ref{sec:pareto}): configurations
are first restricted to $\text{TNR} > 80\%$, a \dRICH-wide post-encoding load below \num{140}~spikes per BC together with a bounded peak load on any single AER chip, and a small Signal+Noise empty-crossing fraction (defined below); the best remaining point on the TPR-versus-latency Pareto front is then adopted.

\begin{figure}[htbp]
  \centering
  \includegraphics[width=1.1\linewidth]{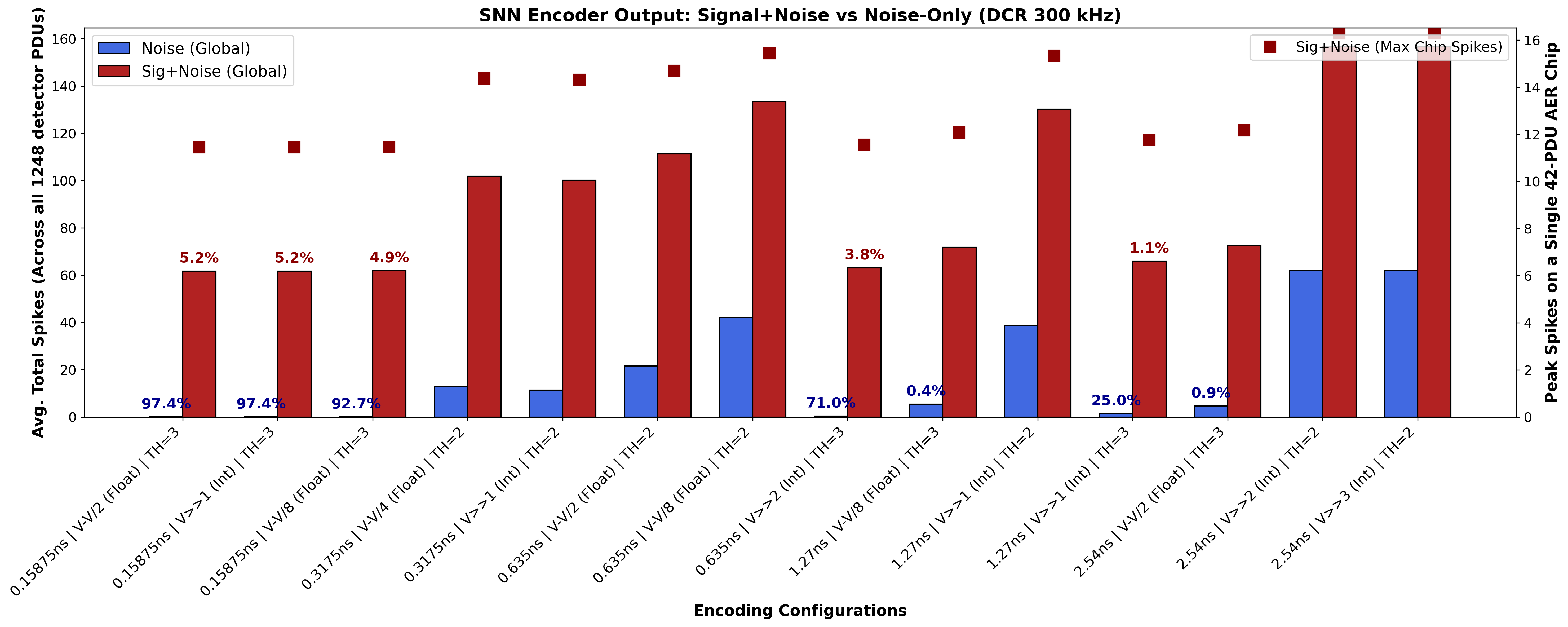}
  \caption{Encoder output spike load for a representative subset of the
  scanned configurations (no early exit), at $\text{DCR}=\SI{300}{\kilo
  \hertz}$. Bars (left axis): average total spikes per BC summed over all
  \num{1248} \dRICH PDUs, for Noise-only (blue) and Signal+Noise (red)
  crossings. Squares (right axis): peak load on a single \num{42}-PDU AER
  chip, which sets the local bus requirement. Percentages annotate, for
  each class, the fraction of bunch crossings that produce \emph{zero}
  encoder spikes; for the Signal+Noise class this is a floor on the
  false-negative rate, since empty crossings convey no information to the
  classifier, while for the Noise-only class a high value reflects
  desirable cost-free rejection. Each configuration is labelled by its
  time bin $\Delta t$, leak rule, and threshold $\theta$.}
  \label{fig:encoder_spikeload}
\end{figure}

The encoder configuration directly sets the spike load presented to the
AER fabric, and hence the cost the serializer and downstream classifier
must absorb. Figure~\ref{fig:encoder_spikeload} characterizes this for a
representative subset of the scanned configurations (no early exit). The
\dRICH-wide average  post-encoder load varies by about an order of magnitude
across the scan. A binding hardware constraint is the peak load on a single \num{42}-PDU AER core
(right axis), which bounds the instantaneous bus occupancy the serializer
must drain within the BC window. A second, classification-side constraint
is the fraction of bunch crossings the encoder leaves \emph{empty}
(percentage annotations in Fig.~\ref{fig:encoder_spikeload}): a crossing
with no output spikes carries no information to the classifier, which can
then only default to the negative (Noise-only) verdict. For the Noise-only
class a large empty fraction is benign - indeed desirable, since those
crossings are correctly rejected at zero AER and compute cost - but for
the Signal+Noise class it sets a hard floor on the achievable
false-negative rate: genuine low-multiplicity signal crossings (dominated
by physics background) that the encoder blanks are irrecoverably lost. The
deployed configuration is therefore chosen to keep the Signal+Noise empty
fraction small while respecting the per-chip load budget; together with
the TPR-versus-latency trade-off of Section~\ref{sec:pareto}, this defines
the space in which the operating point is selected.

For deployment, membrane potentials and synaptic weights are
quantized to fixed-point $\langle b_{\rm int}, b_{\rm frac}\rangle$
representations; the robustness of the classification performance
under this quantization is characterized in Section~\ref{sec:quant}.

\subsection{Pareto frontier of latency versus accuracy}
\label{sec:pareto}

\begin{figure}[htbp]
  \centering
  \includegraphics[width=0.95\linewidth]{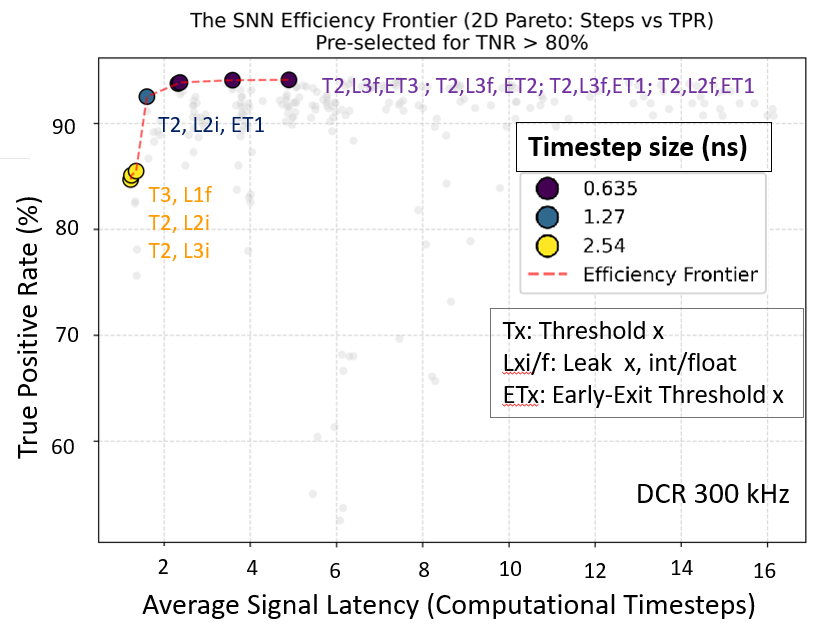}
  \caption{Pareto frontier of TPR versus average signal latency
  (in algorithmic timesteps), over a scan of encoder configurations
  and early-exit thresholds, at fixed $\text{TNR} \ge 80\%$ and
  $\text{DCR} = \SI{300}{\kilo\hertz}$. The chosen operating point
  sits at the knee of the curve.}
  \label{fig:pareto}
\end{figure}

The two free axes of the inference policy --- the encoder
hyperparameters $(\Delta t, k, \theta)$ and the early-exit threshold
$ET$ --- together define a family of operating points trading
classification accuracy for latency. Figure~\ref{fig:pareto} shows
the resulting Pareto frontier in the (average signal latency, TPR)
plane at fixed $\text{TNR} \ge 80\%$. Smaller timestep sizes
($\Delta t = \SI{0.635}{\nano\second}$) consistently outperform larger
ones in TPR, at the cost of more computational timesteps per BC;
a clear knee in the curve sits near $\sim$\num{2} timesteps of
average latency. 

The point we adopt for the current hardware deployment lies in this
knee region: a \SI{1.27}{\nano\second} time bin, a single-place shift
leak ($k=1$), and threshold $\theta=2$ --- the single-bit, same-bin
coincidence-detector limit introduced in Section~\ref{sec:encoder} and
realized by the single-cycle datapath of Section~\ref{sec:encoder_fw}.
That the \emph{most} approximated configuration in the scan also sits at the accuracy/latency knee is a useful outcome of the co-design process: the hardware-simplest operating point and the near-optimal one coincide. 

\subsection{Classification performance versus DCR}

\begin{figure}[htbp]
  \centering
  \includegraphics[width=1\linewidth]{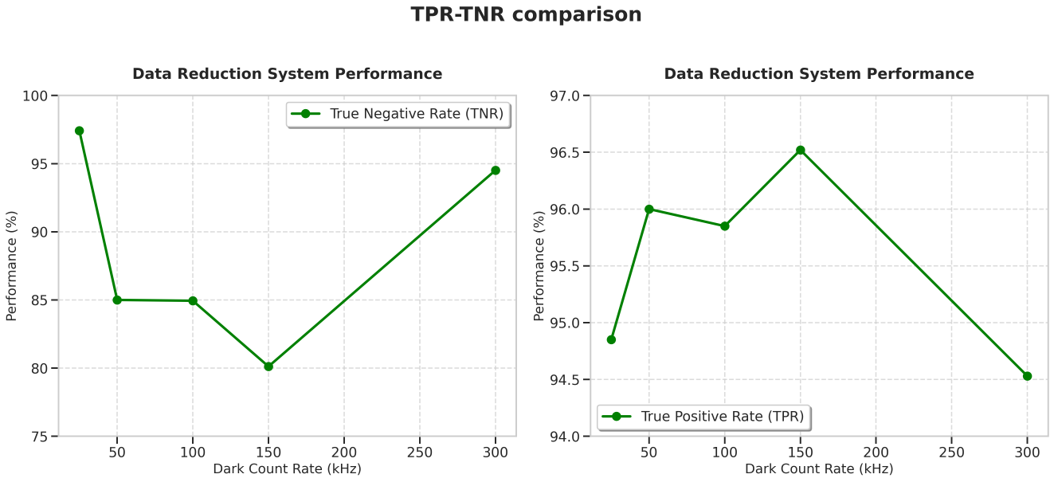}
  \caption{Classification performance of the full system as a function
  of SiPM DCR, at the chosen operating point (time bin
  \SI{1.27}{\nano\second}, encoder threshold $2$, integer leak shift
  $1$). TNR remains above 80\% across the scan and TPR above 94\%.
  Early-exit at $ET=1$ costs approximately three percentage points
  on both metrics.}
  \label{fig:perf_vs_dcr}
\end{figure}

Figure~\ref{fig:perf_vs_dcr} shows the true positive rate (TPR) and
true negative rate (TNR) of the system as a function of DCR, at the
operating point chosen for the hardware deployment (encoder time
bin \SI{1.27}{\nano\second}, threshold $\theta=2$, integer leak shift
$k=1$, no early exit). The TPR remains above 94\% across the full DCR
range, and the TNR meets the $\ge 80\%$ requirement, dipping to its
minimum of $\sim$80\% at \SI{150}{\kilo\hertz}. With early-exit
threshold $ET = 1$, both metrics decrease by approximately three
percentage points; the average classification latency drops from
$T_{\rm max} = 10$ timesteps to $\sim 2$ timesteps, with the
99th-percentile latency at $5$ timesteps. The 99th-percentile figure
sets the point at which SNN execution can be safely truncated under
early exit.

\subsection{Robustness to fixed-point quantization}
\label{sec:quant}

\begin{figure}[htbp]
  \centering
  \includegraphics[width=1\linewidth]{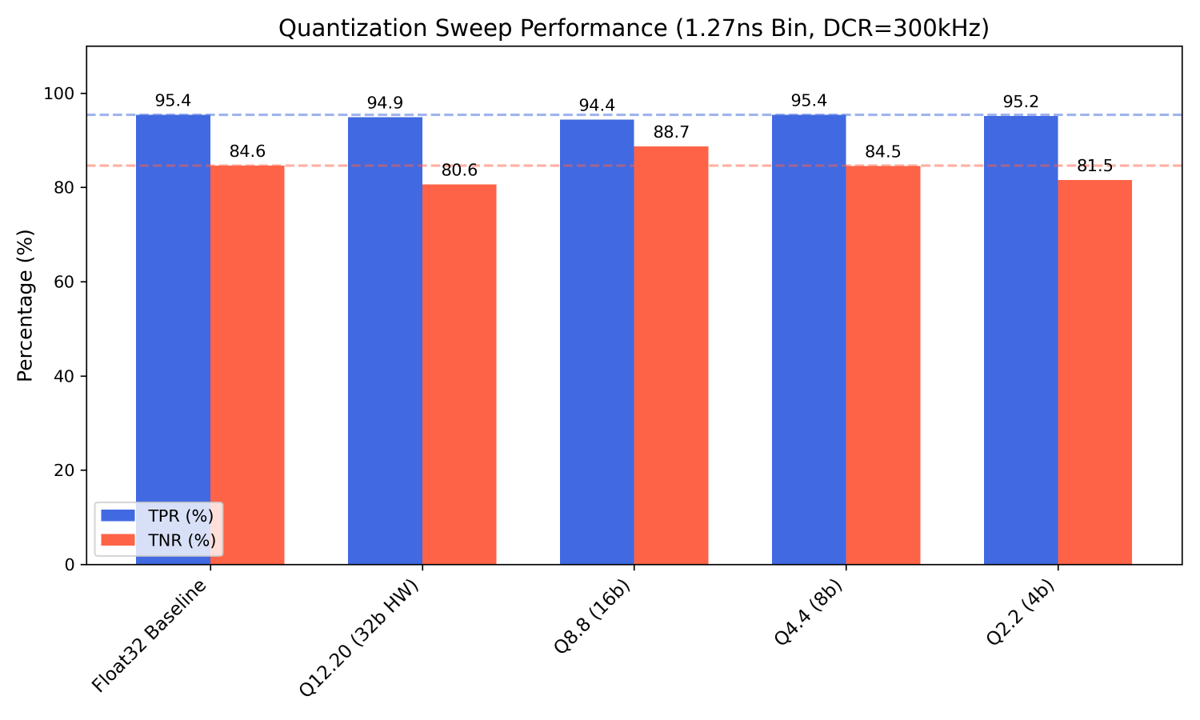}
  \caption{Classification performance under fixed-point quantization
  at the deployed operating point (\SI{1.27}{\nano\second} time bin)
  and $\text{DCR}=\SI{300}{\kilo\hertz}$: TPR (blue) and TNR (red)
  from the floating-point baseline down to the 4-bit Q2.2
  representation. Dashed lines mark the baseline values.}
  \label{fig:quant}
\end{figure}

For hardware deployment the membrane potentials and synaptic weights
are held in fixed-point $\langle b_{\rm int}, b_{\rm frac}\rangle$
(Q-format) arithmetic, so the pipeline must remain accurate under
aggressive quantization. Figure~\ref{fig:quant} shows the TPR and TNR
of the full encoder + classifier pipeline, evaluated at the deployed
operating point and $\text{DCR}=\SI{300}{\kilo\hertz}$, from a
floating-point baseline down to a 4-bit Q2.2 representation. The TPR
stays within one percentage point of the floating-point baseline across
the entire sweep, including at 4 bits; the TNR fluctuates within a few
percentage points of the baseline, with no systematic degradation with
decreasing width.

Two features of the architecture underlie this tolerance. The encoder
is, by construction, the extreme case: its deployed configuration
already runs on a single-bit membrane (Section~\ref{sec:encoder}),
leaving no encoder accuracy to lose to quantization. The downstream LIF
classifier inherits the same robustness --- its short inference horizon
of a handful of timesteps limits the accumulation of rounding error.
This few-bit sufficiency is what keeps the per-neuron FPGA cost
(Section~\ref{sec:hw}) low, and it closes the co-design loop opened in
Section~\ref{sec:encoder}: the same approximations that make the
hardware cheap are precisely the ones the task tolerates. The hardware
testbed of Section~\ref{sec:testbed} currently instantiates a
conservative Q12.20 representation; migrating to the few-bit formats
validated here is part of the planned resource optimization of the
design.

\section{FPGA implementation and hardware characterization}
\label{sec:fpga}

\begin{figure}[hbtp]
  \centering
  \includegraphics[width=.70\linewidth]{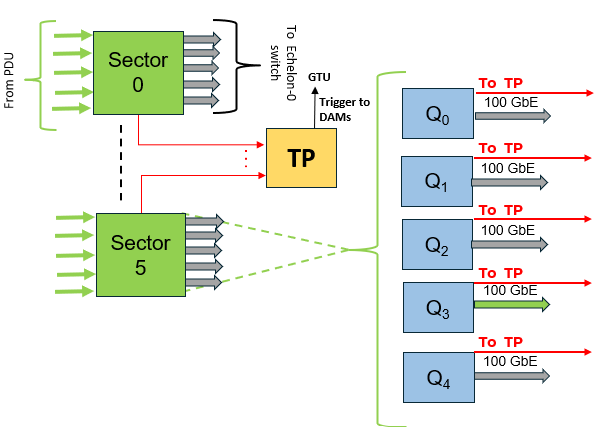}
  \caption{Schematic view of the dRICH online data reduction system deployed on the DAM boards. The green arrows denote the standard RDO data stream directed to the ePIC DAQ switch, while the red arrows indicate the flow of features extracted by each sub-sector SNN. The black arrows represent the interface with the ePIC GTU.}
  \label{fig:dam1}
\end{figure}

\begin{figure}[hbtp]
  \centering

  \begin{minipage}{0.45\linewidth}
    \centering
    
    \begin{subfigure}{\linewidth}
      \centering
      \includegraphics[width=\linewidth]{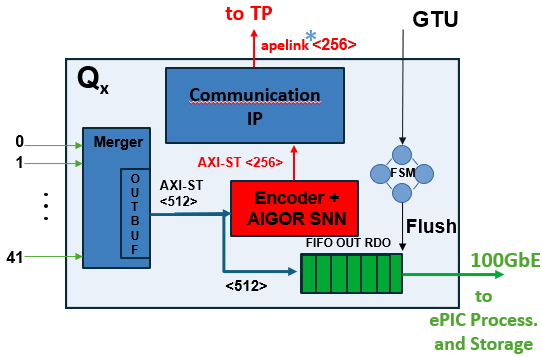}
      \caption{DAM scheme}
      \label{fig:dammlp}
    \end{subfigure}
    
    \vspace{0.45cm}
    
    \begin{subfigure}{\linewidth}
      \centering
      \includegraphics[width=\linewidth]{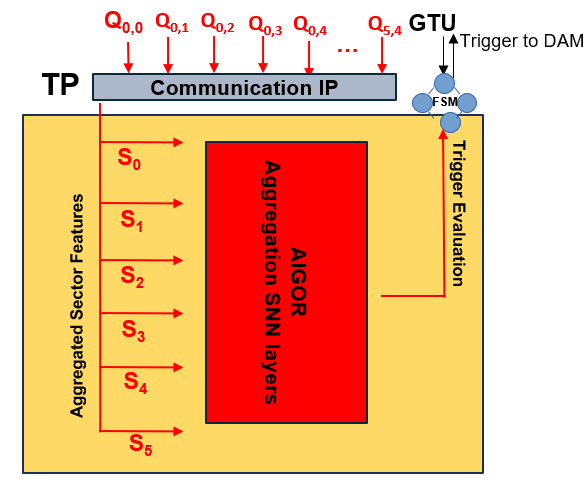}
      \caption{TP scheme}
      \label{fig:tp}
    \end{subfigure}

  \end{minipage}
  \hfill
  \begin{minipage}{0.45\linewidth}
    \centering
    \begin{subfigure}{\linewidth}
      \centering
      \includegraphics[width=\linewidth]{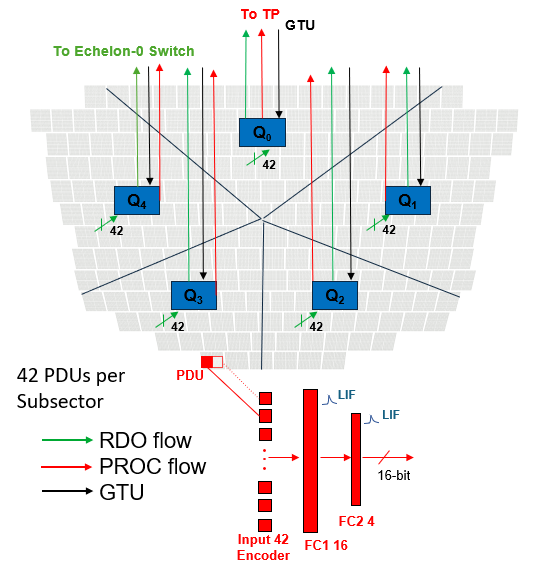}
      \caption{dRICH layout}
      \label{fig:subsec}
    \end{subfigure}
  \end{minipage}

  \caption{Overview of a single dRICH sector: each DAM handles 42 data streams from the sub-sector PDUs (green arrows) and performs the Spike Encoding + SNN inference. The features extracted by each sub-sector SNN are streamed to the TP board via dedicated channels (red arrows), where the final classification is performed. The classification result is then sent back to the DAM through the GTU system (black arrows) and used to enable or inhibit the readout (RDO) data flow to the ePIC Echelon~0 switch.}
  
  \label{fig:dam_tp_third}
\end{figure}

The pipeline is implemented on AMD Versal Premium FPGAs (VPK180 for
the present testbed; FELIX-155 for the final deployment). The mapping
follows the physical readout. Each DAM hosts three major firmware
blocks contributing to the data reduction pipeline: the LIF encoder
for its \num{42} PDU streams, the Sub-sector SNN running on a set of
AIGOR cores, and the event-buffering FIFO that holds the raw RDO data
pending the trigger decision. The Trigger Processor, implemented on an
additional FELIX-155 card, hosts the Aggregation SNN and emits the
trigger verdict via the \ePIC Global Timing Unit (GTU) back to the
DAMs. On positive verdicts the DAMs forward the buffered event
fragments to Echelon-0; on negative verdicts the fragments are flushed.
Inter-FPGA communication of the sub-sector feature vectors uses the
APEIRON IP~\cite{APEIRON}, configured for a 256-bit AER-style payload
per DAM-to-TP link. The detailed firmware implementation over the dRICH backend is shown in
Figure~\ref{fig:dam1} and Figure~\ref{fig:dam_tp_third}. The two
firmware blocks specific to this work are described first --- the
per-PDU encoder pipeline (Section~\ref{sec:encoder_fw}) and the
deployment of the SNN classifier on AIGOR
(Section~\ref{sec:aigor_runtime}) --- followed by the hardware testbed
and its characterization (Sections~\ref{sec:testbed}
and~\ref{sec:hw}).



\subsection{Per-PDU encoder pipeline}
\label{sec:encoder_fw}

\begin{figure}[htbp]
  \centering
  \makebox[\linewidth][c]{%
    \hspace{-0.1cm}%
    \includegraphics[width=1.25\linewidth]{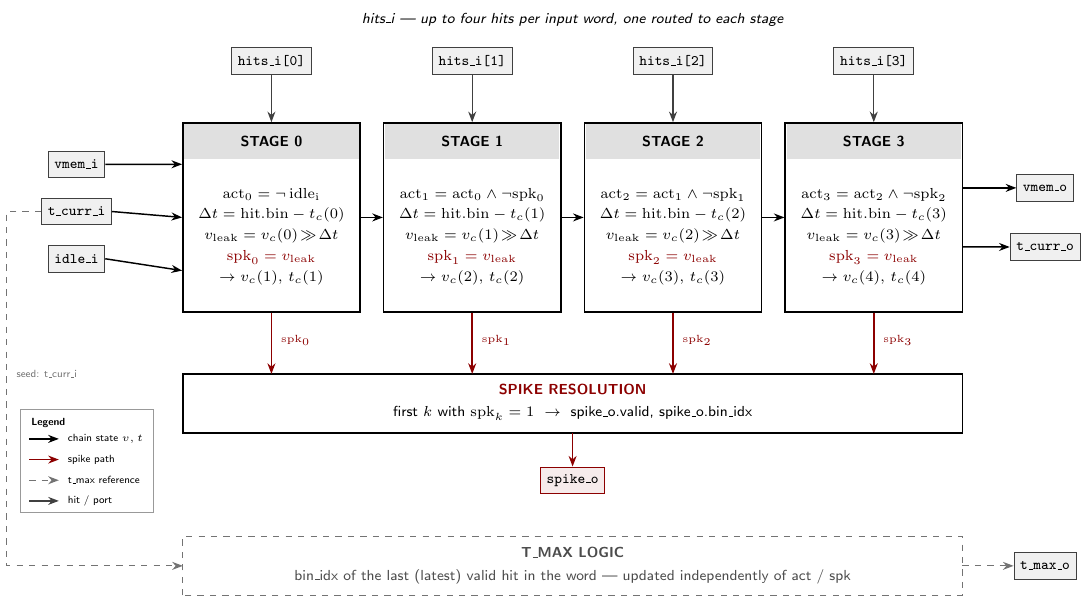}%
  }
  \caption{Per-PDU encoder datapath. The up to four hits of an input
  word traverse four cascaded LIF update stages within a single clock
  cycle; the chain state $(v_c, t_c)$ is seeded from the registered
  per-PDU state and written back at the end of the word. The
  spike-resolution block selects the first stage that fired and tags
  the output spike with its bin index; the $t_{\max}$ logic tracks
  the time of the latest valid hit independently of spiking.}
  \label{fig:encoder_cascade}
\end{figure}

The encoder is the firmware block that resolves the front-end
data-format mismatch identified in Section~\ref{sec:snns}. Hits
arrive from the RDO in dense words of up to four hits each, every
hit carrying a sub-BC bin index. A naive implementation would
serialize each word into the AER bus over four clock cycles ---
unacceptable at the rates we target. We instead exploit the
\textit{single-spike-per-BC} assumption to collapse the per-word
processing into a single clock cycle by cascading four LIF update
stages in combinational logic; the spike-resolution and time-tracking
sub-blocks fold into the same cycle. The architecture is illustrated in
Figure~\ref{fig:encoder_cascade}. The present design is clocked
conservatively at \SI{100}{\mega\hertz} on Versal VPK180 and exploring higher closure is part of the throughput
optimization discussed in Section~\ref{sec:hw}.

Two algorithmic--architectural co-design points are worth
highlighting. First, the cascade works in a single cycle because
\textit{single-spike-per-BC} guarantees that we only ever need to
handle one spike instance per word, so there is no need to serialize
multiple spikes over multiple cycles. Second, a parallel $t_{\max}$
tracker records the bin index of the most recent valid hit in the word,
independent of whether the cascade fired. The leak applied to the
\emph{next} word is anchored to this value, so a word whose stages
stayed idle (e.g. because the membrane never reached threshold) does
not lose track of the running time reference and the per-PDU
integration cannot drift.

\paragraph{Spike serialization.} The at-most-one spike per PDU per BC
must be merged onto the shared AER bus feeding the Sub-sector SNN.
The serializer is a greedy min-tree
arbiter that emits spikes in non-decreasing time-bin order, drops any
late spike below the running watermark, and inserts an end-of-event
token once all \num{42} PDUs have reported. Its cost is bounded by the
number of \emph{spiking} PDUs in the BC: $\le 42$ cycles in the
absolute worst case and $\sim 42\,p_s$ in expectation, where $p_s$ is
the per-PDU per-BC spike probability. At the deployed configuration the
software characterization (Section~\ref{sec:soft}) gives
$p_s \approx 0.055$, i.e. $\sim$\num{2.3} cycles per BC on average ---
which can sit inside the \SI{10}{\nano\second} BC budget running at a clock frequency of 
\SI{250}{\mega\hertz}.

Under the early-exit policy of Section~\ref{sec:earlyexit}, inference
(and hence encoding) need only cover the timesteps up to the decision,
so on early-terminating Signal events the serializer processes fewer
time bins than this per-BC average implies.

\subsection{Classifier deployment on AIGOR}
\label{sec:aigor_runtime}

The Sub-sector and Aggregation SNNs are instantiated on AIGOR, a
modular, event-driven multi-core neuromorphic architecture for
configurable SNN inference on FPGA, developed in our group and
prototyped on Versal Premium VPK180 devices~\cite{AIGOR:wip}. AIGOR is
used here in its feed-forward, streaming configuration; the
architecture itself and its standalone characterization are described
in the companion paper~\cite{AIGOR:wip}.

For this work the network is mapped one-core-per-layer: each LIF layer
of the Sub-sector or Aggregation SNN is hosted by a single AIGOR core,
with strictly feed-forward spike traffic between cores carried as AER
transactions by the AIGOR routing IP, the packet-switched interconnect
of the architecture. Synaptic weights are stored in BRAM; incoming
spikes are routed to the destination workers, which hold per-neuron
state and execute Equation~\ref{eq:LIF}; the outgoing spikes are
arbitrated round-robin onto the next-layer AER bus. The mapping is
strictly feed-forward, with no recurrent spike traffic between cores;
the synchronization cost of the timestep-driven execution model, and
its planned evolution, are discussed in Section~\ref{sec:hw}.

The resource cost of the AIGOR cores hosting the Sub-sector SNN layers
is reported, together with the encoder pipeline, in
Table~\ref{tab:full_resources}: each core occupies well below
\num{1}\% of the available LUTs of a Versal Premium VPK180, leaving
ample headroom for the encoder firmware, the merger, the GTU interface,
and the standard RDO data path that the DAM needs to host.

\subsection{Hardware testbed}
\label{sec:testbed}

\begin{figure}[htbp]
  \centering
  \includegraphics[width=1.05\linewidth]{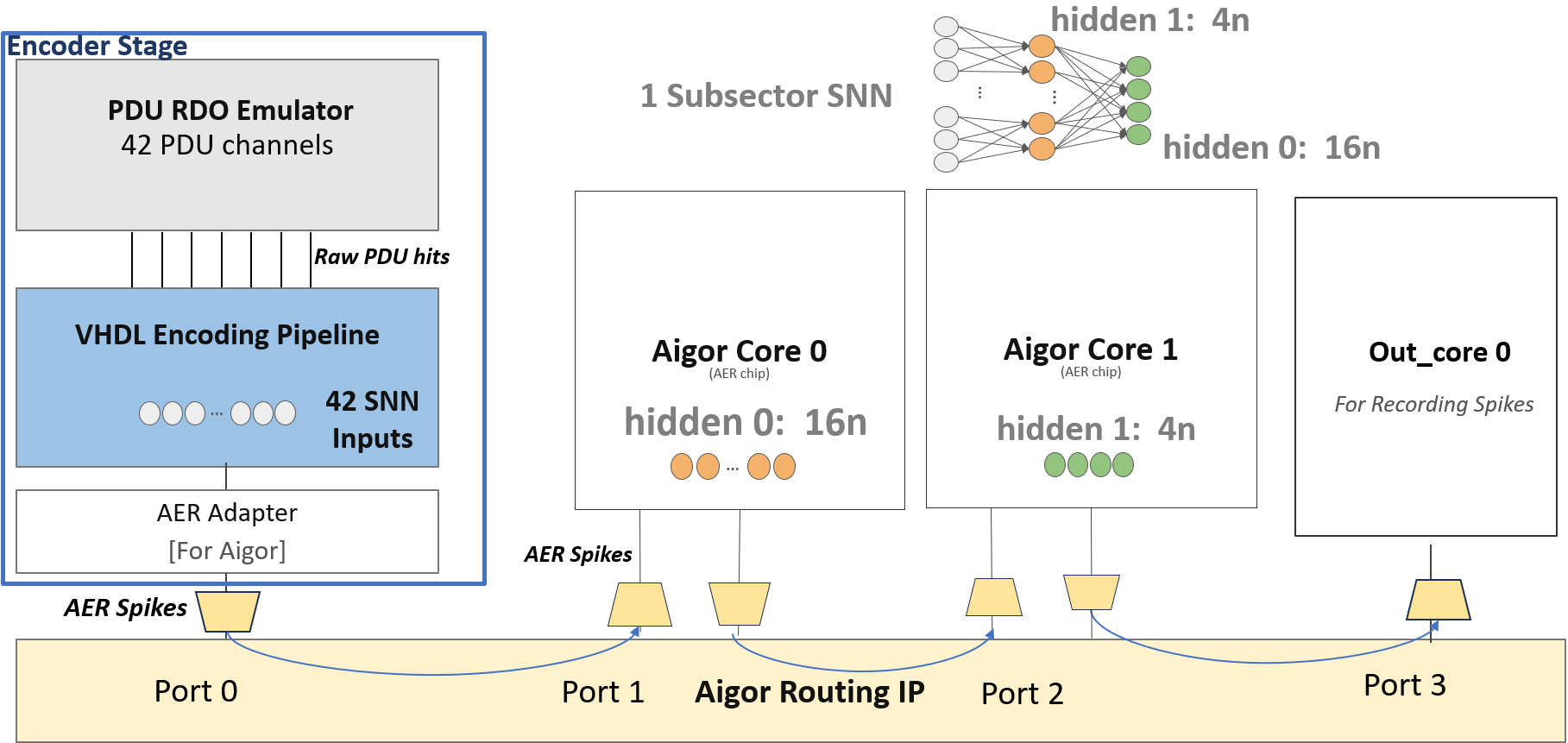}
  \caption{Single-sub-sector hardware testbed on AMD Versal Premium
  VPK180. The PDU/RDO emulator drives the 42-channel VHDL encoding
  pipeline; the encoded AER spike stream enters the AIGOR routing IP
  at Port~0 and is switched to the two AIGOR cores hosting the
  16-neuron and 4-neuron Sub-sector SNN layers and to the out-core
  that records the output spike stream.}
  \label{fig:hw_testbed}
\end{figure}

We have implemented and characterized a single-sub-sector instance of
the pipeline on an AMD Versal Premium VPK180 evaluation board. The
testbed (Figure~\ref{fig:hw_testbed}) consists of a PDU/RDO emulator
that streams synthetic event data reproducing the hit distribution at
\SI{300}{\kilo\hertz} DCR; the VHDL encoding pipeline, containing
the \num{42} per-PDU LIF encoders, a greedy min-tree AER serializer,
and an AER adapter for AIGOR; two AIGOR cores hosting the Sub-sector
SNN's hidden layers (16 LIF neurons in Core~0 and 4 LIF neurons in
Core~1); and a recording out-core that captures the output spike
stream. The blocks communicate through the AIGOR routing IP: the
encoded spike stream enters the routing fabric at Port~0 and is
switched to the two SNN cores and to the out-core, so the testbed
exercises the same packet-switched AER transport that carries
core-to-core traffic in the full multi-FPGA deployment, here confined
to a single device. Inter-FPGA transport is deferred to the full
deployment.

\subsection{Measured performance}
\label{sec:hw}

The encoder pipeline is fully parallel across PDUs: each of the
\num{42} LIF encoders integrates its own SiPM hits independently, and
the greedy min-tree AER serializer multiplexes their outputs onto the
shared bus, with the single-spike-per-BC policy bounding the
serialization cost (Section~\ref{sec:encoder_fw}). On a synthetic
dataset reproducing the per-PDU hit distribution at
\SI{300}{\kilo\hertz} DCR ($\sim$\num{1.4} hits/PDU/BC on
average), the encoder alone sustains a throughput in the range
\SIrange{1}{10}{\mega\hertz}, depending on the per-BC hit count. The
complete encoder + Sub-sector SNN chain, run on a semi-realistic input
trace (one Signal+Noise crossing per \num{100} Noise-Only crossings,
as in the \ePIC simulation), sustains a measured
$\sim$\SI{1.7}{\mega\hertz}, validating the complete pipeline ---
from raw hit words to classifier output spikes --- in hardware.

Future work will aim to progressively close the gap to the nominal
\SI{100}{\mega\hertz} bunch-crossing rate. Four optimizations have
been identified with the largest expected impact:
\textbf{multi-spike AER packets}, allowing one AER transaction to
carry several simultaneous spikes from the same core at the price of
parallel access to the synaptic memory;
\textbf{event-driven execution}, evolving the runtime dynamics from
timestep-driven toward event-driven operation, so that algorithmic
time no longer needs to be marked explicitly on every link at every
timestep --- fusing the synchronization information into the spike
words themselves being a first step in this direction;
\textbf{clock frequency} above the conservative
\SI{100}{\mega\hertz} used for the present measurements; and
\textbf{early-exit truncation}, which reduces the per-event
computational cost in proportion to the timestep savings reported in
Section~\ref{sec:earlyexit}.
Quantifying the individual contribution of each optimization is part
of this effort.

\paragraph{Resource utilization.} The complete single-sub-sector
pipeline (Table~\ref{tab:full_resources}) occupies only 1.8\% of the
available LUTs, 0.72\% of the registers, 1 DSP slice, and 1.75\% of
the BRAM resources of the VPK180 device, leaving substantial margin for
the DAM firmware managing communication with the rest of the DAQ chain
and for the readout buffer.

\begin{table}[htbp]
  \centering
  \caption{Resource utilization of the full single-sub-sector pipeline
  (PDU emulator excluded) on AMD Versal Premium VPK180.}
  \label{tab:full_resources}
  \small
  \setlength{\tabcolsep}{5pt}
  \begin{tabular}{lcccc}
    \toprule
    Block & LUT (\%) & FF (\%) & DSP (\%) & BRAM (\%) \\
    \midrule
    LIF encoder + AER serializer & 16.9k (0.50) & 22.6k (0.34) & 1 (0.007) & 86.0 (1.74) \\
    AIGOR core 0 (16-LIF layer)  & 23.1k (0.69) & 11.3k (0.17) & 0 (0.00) & 0.0 (0.00) \\
    AIGOR core 1 (4-LIF layer)   & 20.6k (0.61) & 14.7k (0.22) & 0 (0.00) & 0.5 (0.01) \\
    \midrule
    \textbf{Total} & \textbf{60.6k (1.80)} & \textbf{48.5k (0.72)} &
                     \textbf{1 (0.007)} & \textbf{86.5 (1.75)} \\
    \bottomrule
  \end{tabular}
\end{table}

\paragraph{End-to-end latency.} Beyond throughput, the latency from a
PDU hit to the emitted trigger verdict sets the buffering depth the DAM
must provide. In the present testbed it is bounded by three
contributions: the single-cycle per-PDU encoder
(Section~\ref{sec:encoder_fw}), the AER serialization, which costs at
most \num{42} cycles in the worst case and $\sim$\num{2.3} on
average at the deployed spike probability, and the SNN inference
horizon set by the early-exit timestep budget
(Section~\ref{sec:earlyexit}). Alongside the throughput optimizations
above, the characterization campaign of the ongoing work includes a
measured per-stage latency decomposition and an energy-per-inference
measurement --- the metric on which the event-driven compute pattern,
active only on the sparse spike load of a crossing and largely silent
on the dominant Noise-Only class, is expected to pay off most.

\section{Discussion}
\label{sec:disc}

The largest single contribution to the data reduction comes from the
LIF temporal-coincidence encoder, which discards more than 90\% of the
raw hit volume before any learned classifier is applied, at the cost of
one specialized LIF neuron per PDU with a state of a few bits. This
pattern --- a small temporal filter between the front end and the
learned classifier, with its hyperparameters selected jointly with the
downstream network --- may transfer to other detectors whose signal is
carried by fine timing structure over an approximately uniform noise
background. The compactness of the encoder also makes deployment closer
to the readout electronics, upstream of the RDO--DAM optical links, an
attractive longer-term direction for event-driven front-end processing,
although its integration in the existing front-end electronics has not
been assessed here and would raise its own design questions.

On the hardware side, the measured \SI{1.7}{\mega\hertz} is
dominated by the timestep-synchronization traffic of the inference
fabric: the timestep-based dynamics of the classifier require the
advance of algorithmic time to be marked explicitly on every AER link,
and on the sparse post-encoder spike streams these synchronization
words can outnumber the spikes themselves. Reducing this cost requires evolving the execution model of the
inference fabric from timestep-driven toward event-driven dynamics;
together with multi-spike AER packets, higher clock frequency, and
early-exit truncation (Section~\ref{sec:hw}), this defines the
optimization path being pursued toward the \SI{100}{\mega\hertz}
bunch-crossing rate.
\section{Conclusion}
\label{sec:conc}

We have presented an online data reduction pipeline for the \ePIC
\dRICH detector based on spiking neural networks. The pipeline is
structured in two stages: a per-PDU LIF temporal-coincidence encoder
that achieves $>$90\% data sparsification by exploiting the
$\sim$\SI{2}{\nano\second} burst structure of Cherenkov hits against a
uniform DCR background, and a distributed two-tier SNN classifier
that emits the Signal+Noise versus Noise-Only verdict. On simulated
\ePIC events the system reaches $\text{TPR} > 94\%$ at $\text{TNR}
\ge 80\%$ across the operational DCR range, and an early-exit
inference strategy reduces the average classification latency to
$\sim$\num{2} algorithmic timesteps. A single-sub-sector hardware
testbed on AMD Versal Premium FPGAs, integrating the LIF encoder
with the AIGOR multi-core neuromorphic architecture, validates the
design at a measured $\sim$\SI{1.7}{\mega\hertz} on hardware;
future work will pursue the identified optimizations toward the
\SI{100}{\mega\hertz} bunch-crossing rate. The technique, a
small, hardware-friendly LIF encoder coupled to a distributed SNN
classifier, operating directly at the front-end boundary of a
streaming readout chain, may be relevant to other timing-driven
detector applications operating at high rate.

\section*{Acknowledgments}
P.~Perticaroli is a PhD student enrolled in the National PhD programn in Artificial Intelligence, XXXIX cycle, course on Health and Life Sciences, organized by Università Campus Bio-Medico di Roma. The authors acknowledge the financial support of the Istituto Nazionale di Fisica Nucleare (INFN), Commissione Scientifica Nazionale 5 (CSN5), through the BRAINSTAIN project.


\end{document}